\documentclass[pss]{wiley2sp} 
\usepackage{amsmath}
\usepackage{bm}              
\usepackage{w-greek}         
\usepackage{graphicx,psfrag}
\usepackage{amsfonts}
\usepackage{amssymb}
\usepackage{amsmath}
\usepackage{epsfig, color}
\usepackage{setspace}
\usepackage{epstopdf}

\def\vk{{\vec k}}

\def\vd{{\vec d}}
\def\vq{{\vec q}}

\def\vk{{\vec k}}

\def\vd{{\vec d}}
\def\vn{{\vec n}}
\def\vK{{\vec K}}

\def\vV{{\vec V}}
\def\vA{{\vec A}}

\def\ve{{\vec e}}

\def\ve{{\vec e}}

\mathchardef\sigma="711B

\def\vd{{\vec d}}
\def\ve{{\vec e}}

\def\nhat{{\hat n}}

\def\ve{{\vec e}}

\def\ve{{\vec e}}

\def\nhat{{\hat {\vec n}}}
\def\dhat{{\hat {\vec d}}}

\def\ve{{\vec e}}

\begin{document}

\title{Floquet topological insulators}

\titlerunning{Floquet topological insulators }

\author{%
  J\'er\^ome Cayssol\textsuperscript{\Ast,\textsf{\bfseries 1,2}},
  Bal\'azs D\'ora\textsuperscript{\textsf{\bfseries 3,4}},
  Ferenc Simon\textsuperscript{\textsf{\bfseries 4}},
  Roderich Moessner\textsuperscript{\textsf{\bfseries 1}}}

\authorrunning{J. Cayssol et al.}

\mail{e-mail:
  \textsf{jcayssol@pks.mpg.de}}

\institute{%
   \textsuperscript{1}\,Max-Planck-Institut f\"ur Physik komplexer Systeme, N\"othnitzer Str. 38, 01187 Dresden, Germany\\
  \textsuperscript{2}\,LOMA (UMR-5798), CNRS and University Bordeaux 1, F-33045 Talence, France\\
  \textsuperscript{3}\,BME-MTA Exotic  Quantum  Phases Research Group, Budapest University of Technology and Economics, Budapest, Hungary\\
  \textsuperscript{3}\,Department of Physics, Budapest University of Technology and  Economics, Budafoki \'ut 8, 1111 Budapest, Hungary
}

\received{XXXX, revised XXXX, accepted XXXX} 
\published{XXXX} 


\abstract{%
%
%
%
\abstcol{%
 Topological insulators represent unique phases of matter with insulating bulk and conducting edge or surface states, immune to small perturbations such as backscattering due to disorder. This stems from their peculiar band structure, which provides topological protections. While conventional tools (pressure, doping etc.) to modify the band structure are available, time periodic perturbations can provide tunability by adding time as an extra dimension enhanced to the problem. 
  }{%
  In this short review, we outline the recent research on topological insulators in non equilibrium situations. Firstly, we introduce briefly the Floquet formalism that allows to describe steady states of the electronic system with an effective time-independent Hamiltonian. Secondly, we summarize recent theoretical work on how light irradiation drives semi-metallic graphene or a trivial semiconducting system into a topological phase. Finally, we show how photons can be used to probe topological edge or surface states.  }}

\maketitle   

\section{Introduction}
Topological phases, including topological insulators (TIs) \cite{Kane:2005a,Kane:2005b,Bernevig:2006,KaneRMP:2011,QiRMP:2011} and Chern insulators (CIs) \cite{Haldane:1988,Qi:2006,Tang:2011,Sun:2011,Neupert:2011}, represent unique states of matter owing to the robust, topological protection of their conducting edge or surface states. CIs (also called Quantum Anomalous Hall phases) appear in lattice models with bands carrying a finite Chern number, thereby realizing a charge Quantum Hall effect in the absence of net magnetic flux \cite{Haldane:1988,Qi:2006}. The two dimensional TI, namely the quantum spin Hall (QSH) state, has been predicted for a variety of systems including
graphene \cite{Kane:2005a,Kane:2005b}, inverted HgTe/CdTe quantum wells \cite{Bernevig:2006}, exotic lattice models \cite{Weeks:2010,guo,sun} and multi-component ultracold fermionic atoms in appropriate optical lattices \cite{Jiang:2011,stanescu,goldman}. In all proposals for CIs, the chiral edge state(s) originate(s) from a non-trivial background gauge field \cite{Haldane:1988,Qi:2006,Tang:2011,Sun:2011,Neupert:2011}. The gapless helical edge state requires a subtle band inversion driven by spin-orbit coupling \cite{KaneRMP:2011,QiPhysToday:2010}. Such topologically non-trivial band structures are rather scarce since they require careful Bloch band structure engineering \cite{Haldane:1988,Qi:2006,Tang:2011,Sun:2011,Neupert:2011,Kane:2005a,Kane:2005b,Bernevig:2006} and a high degree of sample control \cite{Konig:2007,Roth:2009}. 

Bloch states and energy bands arise from spatially periodic Hamiltonians in condensed matter systems. Extending the periodicity 
in the time domain by applying a time-periodic perturbation increases the tunability of the Hamiltonian since the  temporal analogue of Bloch states (the Floquet states) can be manipulated via the polarization, periodicity and amplitude of the external perturbation. Recently, topological phases of periodically driven quantum systems have been characterized \cite{kitagawa} using Floquet theory, thereby 
extending the general classification of topological phases at equilibrium \cite{Ryu:2008,Hughes:2008,Kitaev:2009}. Interestingly, novel topological edge states can be induced by shining  electromagnetic radiation on a topologically trivial insulator, e.g. a 
non-inverted HgTe/CdTe quantum well with no edge state in the static limit \cite{lindner}. It is also possible to open gaps at the Dirac point of graphene and even drive graphene into the topological Haldane 
phase by simple irradiation with suitably chosen parameters \cite{kitagawa,Oka:2009,Calvo:2011,Kitagawa:2011,Morell:2012}. Therefore DC transport is expected to be drastically modified under such irradiation \cite{Kitagawa:2011,GuPRL:2011}. 

Besides, a time-dependent perturbation may also be harmful to the coherence of the edge/surface states of TIs by introducing dissipation and breaking time-reversal symmetry (maintaining the latter being an essential condition for the existence of helical edge states of TIs). It is therefore natural to investigate to what extent the steady state of a TI remains robust against time-dependent perturbations and how its electrical and magnetic properties are altered. For weak amplitudes, the external time-dependent field can also be used as a probe of the topological character of the edge state.

Here, we consider only the effect of a classical time-periodic perturbation, for instance the electromagnetic wave generated by a laser or a RF-source. We first review the Floquet formalism in the context of CIs and TIs and introduce the corresponding topological invariants (Sec. \ref{sectionFormalism}). Then in Sec. \ref{sectionCreation} we describe how non-trivial topological phases can be created by shining light on a trivial insulator or on semi-metallic graphene. Sec. \ref{sectionQSH} describes our work on the photocurrent generated by shining an electromagnetic wave onto the helical edge states of a QSH insulator \cite{Dora:2012}. Finally Sec. \ref{sectionPhotocurrent2D} reviews related works on the photocurrent generation in the surface states of 3D topological insulators \cite{Hosur:2011,Ueda:2012}.   

\section{Floquet formalism \label{sectionFormalism}}

Here we review the aspects of Floquet formalism that are useful to grasp the physics of topological insulators in presence of 
external periodic driving. More general presentations of the Floquet formalism are available  \cite{sambe,dittrich}. 
\subsection{Effective Hamiltonian}
We start from the time-independent Bloch Hamiltonian of a two-band insulator:
\begin{equation}
\mathcal{H}_0(\vk)=\epsilon_{0}(\vk)\, {\bold I}_{\rm 2x2}+  \vd(\vk) \cdot \bm{\sigma} ,
\label{Hamilton2by2Flo}
\end{equation}
which describes the system in the absence of irradiation. The vector of Pauli matrix $\bm{\sigma}=(\sigma_x,\sigma_y,\sigma_z)$ represents an isopin degree of freedom that can be the real electronic spin, sublattice index, orbital index etc., and that couples with the orbital motion. In the following, this Hamiltonian can be either the full Hamiltonian (see paragraph \ref{subsectionGraphene}), or alternatively it can be considered as one block of a 4-band time-reversal invariant insulator, the other block then being $\mathcal{H}^{*}(-\vk)$ (see paragraph \ref{subsectionHgTe}). 

We now consider the time-dependent perturbation:
\begin{equation}
\mathcal{V}(\vk,t)= \vV(\vk) \cdot \bm{\sigma} \cos (\omega t),
\label{Hamilton2by2TimeDepLinear}
\end{equation}
which describes coupling to a linearly polarized monochromatic electromagnetic wave, $\vV(\vk)$ being a time-independent vector. Other polarizations are also possible and may lead to different behaviors in terms of gap openings and/or topological properties. For instance, the coupling to a classical circularly polarized monochromatic wave can
be written as:
\begin{equation}
\mathcal{V}(\vk,t)=V(\vk) \left( \sigma_{-} \exp (i \omega t) + \sigma_{+} \exp (-i \omega t) \right),
\label{Hamilton2by2TimeDepCirc}
\end{equation}
where $\sigma_{\pm}=\sigma_x \pm i \sigma_y$, and $V(\vk)$ is a time-independent real number.

Under irradiation, the quantum states evolve as $\Psi(t)=\mathcal{U}(t,t_0) \Psi(t_0)$ where the evolution operator is given by:
\begin{equation}
\mathcal{U}_{\vk}(t,t_0)=\mathcal{T}_t \exp \left( -i \int_{t_0}^{t} dt' \mathcal{H}(\vk,t') \right),
\label{EvolutionOp}
\end{equation}
$\mathcal{T}_t$ being the time-ordering operator and $\mathcal{H}(\vk,t)=\mathcal{H}_0(\vk) + \mathcal{V}(\vk,t)$ the full time-dependent Hamiltonian. The Floquet Hamiltonian of the crystal is defined as the stationary Bloch Hamiltonian that would yield the same unitary evolution after one period ($T$) 
of the driving field:
\begin{equation}
e^{-i H_{F}(\vk) T  } =  \mathcal{U}_{\vk}(T+t_0,t_0).
\label{HeffectiveDefinition}
\end{equation}
Hence the Floquet Hamiltonian $H_F$ is a stationary effective Hamiltonian that describes the "stroboscopic" evolution of the system
after each period of the driving field. This effective Floquet approach is valid as long as the period $T$ of the driving field is the shortest time scale in the system, a condition that might
be easier to realize in cold atom systems than in solid state electronic systems. The Floquet Hamiltonian, however, does not contain all information about the topological properties of our system, for example it cannot reveal the Z$\times$Z or Z$_2 \times$Z$_2$ topological invariants, as shown in Refs \cite{Jiang:2011,Asboth:2012}.  

\subsection{Floquet Chern number}
Being also a $2 \times 2$ matrix the effective Floquet Hamiltonian can be parametrized as:
\begin{equation}
\mathcal{H}_{F}(\vk)=\epsilon_{F}(\vk)\, {\bold I}_{\rm 2x2}+ \vn(\vk)  \cdot \bm{\sigma},
\label{Floquet2by2}
\end{equation}
where the new functions $\epsilon_{F}(\vk)$ and $ \vn(\vk)$ are computed from Eq. (\ref{HeffectiveDefinition}). This Hamiltonian can be studied using the methods suitable to classify insulators at equilibrium \cite{Thouless1982,Qi:2006}. 
In particular a Chern number can be defined as the winding number of the mapping $\vk \rightarrow \bm{\nhat}(\vk)=\vn(\vk)/|\vn(\vk)|$ from the torus $T^2$ towards the unit sphere $S^2$:
\begin{equation}
C_F =\frac{1}{4 \pi}  \int d^2 \vk   \left(  \frac{ \partial \bm{\nhat}(\vk)}{\partial k_x}     \times  \frac{ \partial \bm{\nhat}(\vk)}{\partial k_y}   \right) \cdot   \bm{\nhat}(\vk),
\label{WindingFloquet}
\end{equation}
provided one has $|\vn(\vk)| \neq 0$ in the whole Brillouin zone, namely the Floquet effective Hamiltonian has a gapped spectrum. The number $C_F$ can be different from the Chern number 
\begin{equation}
C_I =\frac{1}{4 \pi}  \int d^2 \vk   \left(  \frac{ \partial \bm{\dhat}(\vk)}{\partial k_x}     \times  \frac{ \partial \bm{\dhat}(\vk)}{\partial k_y}   \right) \cdot   \bm{\dhat}(\vk),
\label{WindingInitial}
\end{equation}
which describes the topology of the Bloch wave functions for the non irradiated insulator. Recently it has been demonstrated that the Floquet Chern number may be non zero ($C_F \neq 0$) even if the non irradiated system is an ordinary band insulator with no edge states and zero Chern number ($C_I=0$). This means that the effective Floquet Hamiltonian also has a chiral edge state, as we discuss next.




\section{Creating new topological phases by light \label{sectionCreation}}
Here we start from semi-metallic graphene ($C_I$ not defined) or from an insulator with trivial topology ($C_I=0$). The general idea is to investigate whether a non equilibrium 
perturbation can drive such systems into some topological phase, characterized by a nonzero topological invariant ($C_F \neq 0$) and whether there are transport signatures to such a light induced topological phase transition. Using Floquet theory, different groups have answered positively to these questions. We shall consider separately the cases of graphene and other semiconducting systems (HgTe/CdTe or Rashba coupled two dimensional electron gas) for clarity. 

\subsection{Graphene \label{subsectionGraphene}}

Opening gaps in graphene is a very important issue both for transistor applications (realization of a non conducting off-state, confinement of Dirac carriers into narrow channels etc.) and for fundamental physics (realization of Haldane phase). Unfortunately (or fortunately) the Dirac points are very robust since they are protected by fundamental symmetries like space-inversion and time-reversal, while spin-orbit coupling, which is allowed by symmetry to open a gap, is too weak.  

\medskip

{\it Photoinduced gaps.} In spite of this robustness, it has been predicted that circularly polarized light can open a gap at the Dirac points \cite{Oka:2009}. Graphene is described by the time-dependent Bloch Hamiltonian:
\begin{equation}
\mathcal{H}(\vk,t)=v_F \left(  \sigma_x \tau_z (q_x +e A_x(t))+ \sigma_y  (q_y + eA_y(t)) \right),
\end{equation}
now including the coupling to the electromagnetic field. The two-dimensional momentum $\vq= \vk - \xi \vK=q_x \ve_x + q_y \ve_y$ is measured from the Dirac points locations $\xi \vK$ ($\xi=\pm 1$). The Pauli matrices $\sigma_z$ and $\tau_z$ corresponds to the sublattice isospin and the valley index, respectively. On the sample, the vector potential $\vA(t)=(A_x(t),A_y(t))$ is taken to be:
\begin{equation}
\vA= - {A_0} \left( \sin(\omega t)  \ve_x + \sin(\omega t - \phi ) \ve_y \right),
\end{equation}
where the angle $\phi$ allows to tune the polarization of the wave and $E_0 = A_0 \omega$ is the field amplitude. We assume that the sample is smaller than the wave length of light or that it is irradiated at normal incidence by a plane wave so there is no spatial dependence of the driving field. 

\medskip

\begin{figure}[h!]
{\includegraphics[width=7.0cm]{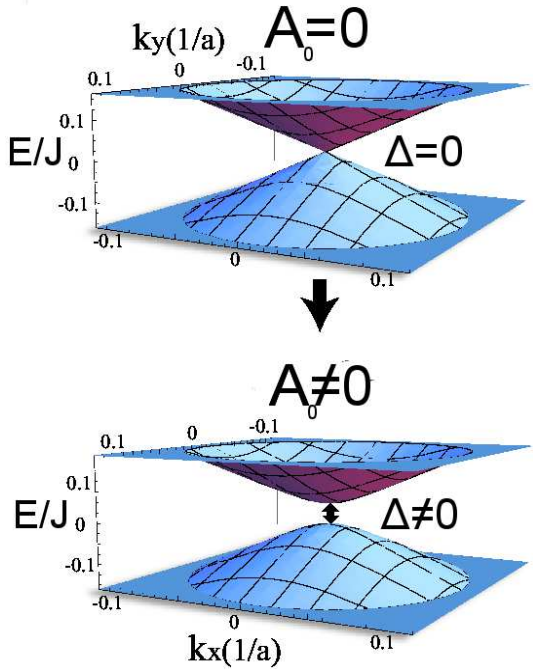}}
\caption{(Color online) The energy spectrum of Dirac electrons  near
one of the Dirac points is shown for $A_0  =0$ (upper figure) and the spectrum of $H_{F}$ for the system under
the application of light with $A_0  \neq 0$ (lower figure), opening a finite gap at the Dirac point. $J_0$ is the hopping amplitude and $a$ the lattice parameter of the graphene honeycomb lattice. Picture taken from Ref. \cite{Kitagawa:2011}.}
\label{kit2}
\end{figure}

For small laser power, i.e. $e A_0 v_F \ll \hbar \omega$, the effective Hamiltonian is well-approximated by  $\mathcal{H}_{F}(\vk) \simeq \mathcal{H}_0 + [H_{-1},H_{+1}]/\hbar \omega $, namely:
\begin{equation}
\mathcal{H}_{F}(\vk) = \mathcal{H}_0 -\frac{(e A_0 v_F)^2}{\hbar \omega} \sin \phi \, \sigma_z \tau_z,
\label{FloquetGraphene}
\end{equation}
where
\begin{equation}
H_{m}=H_{m}(\vk)=\frac{1}{T} \int_{0}^{T} dte^{im \omega t} H(\vk,t),
\end{equation}
is the $m$-th Fourier harmonic of the time-periodic Hamiltonian $H(t)$. Interestingly, the generated mass term is of the Haldane type $\sigma_z \tau_z$, namely it changes sign from one valley to the other \cite{Haldane:1988}. 
This gap opening originates in (off-resonant) virtual photon absorption processes that give a mass to Dirac fermions \cite{Kitagawa:2011}, as shown in Fig. \ref{kit2}.
The amplitude of the photoinduced gap/mass at the Dirac points, 
\begin{equation}
\Delta = 2 \frac{(e A_0 v_F)^2}{\hbar \omega} \sin \phi,
\end{equation}
depends drastically upon the polarization, being maximal for circular polarization ($\phi=\pm \pi/2$) while cancelling out for linear polarization of light ($\phi=0,\pi$) as shown in Ref. \cite{Calvo:2011}. This can be understood in the following way. The circular polarization provides the chirality which is mandatory for the Haldane phase to occur. In contrast, a linear polarization, made of equal superposition of clockwise and anticlockwise circular polarizations, does not break time-reversal symmetry and cannot lead to a Haldane mass term. Finally the gap is given by the formula:
\begin{equation}
\Delta =  16 \pi \alpha \frac{v_F^2 I}{\omega^3} \sin \phi,
\end{equation}
where $I$ is the laser intensity (W/m$^2$) and $\alpha \simeq 1/137$ is the fine structure constant.


{\it DC electrical transport through irradiated graphene.} In the static case, the Haldane mass is associated with a finite Chern number for the occupied band and therefore a quantized Hall conductance $e^2/h$. For graphene under circularly polarized light, it was pointed out that such photoinduced topological mass term should also manifest itself in transport through a Hall effect \cite{Oka:2009,Kitagawa:2011}. Nevertheless, an appropriate formalism needs to be used to compute DC Hall response functions in presence of steady states described as Floquet states \cite{Kitagawa:2011,GuPRL:2011}.  

Finally the Haldane model (starting with a topological mass in the absence of irradiation) has also been studied in presence of time-periodic driving which allows to tune the parameters of the Haldane model \cite{inoue}.

\subsection{HgTe/CdTe heterostructures \label{subsectionHgTe}}

The HgTe/CdTe quantum well is an important system as it is the first one where the QSH state and its edge states have been reported experimentally. Nevertheless the transition between the trivial insulating state (non-inverted regime) and the QSH state (inverted regime) requires to change the width of the well (central layer) of HgTe at the nanoscale. It is clearly desirable to be able to trigger this transition by some external knob like a gate voltage or irradiation with light. Although a system has been proposed and tested that allows control of the band inversion by operating a gate voltage \cite{Liu:2008,Knez:2011}, the idea of using light seems promising especially in view of the potential applications for optronics.  

{\it Transition.} We start from a non-inverted HgTe/CdTe quantum well described by the Hamiltonian: 
\begin{equation}
\mathcal{H}_{4\rm{x}4}(\mathbf{k})=%
\begin{pmatrix}
\mathcal{H}_{0}(\mathbf{k}) & 0 \\ 
0 & \mathcal{H}_{0}^{\ast}(-\mathbf{k})%
\end{pmatrix}%
, 
 \label{BHZ4x4bis}
\end{equation}
where
\begin{equation}
\mathcal{H}_0(\vk)=\epsilon_{0}(\vk)\, {\bold I}_{\rm 2x2}+  \vd(\vk) \cdot \bm{\sigma} ,
\label{Hamilton2by2Flobis}
\end{equation}
and the vector:
\begin{equation}
\vd(\vk) =(A k_x,A k_y,M-B\vk^2)
\label{Hamilton2x2TBAk}
\end{equation}
parametrizes an effective spin-orbit coupling near the center of the first Brillouin zone (FBZ).

\begin{figure}[h!]
{\includegraphics[width=7.0cm]{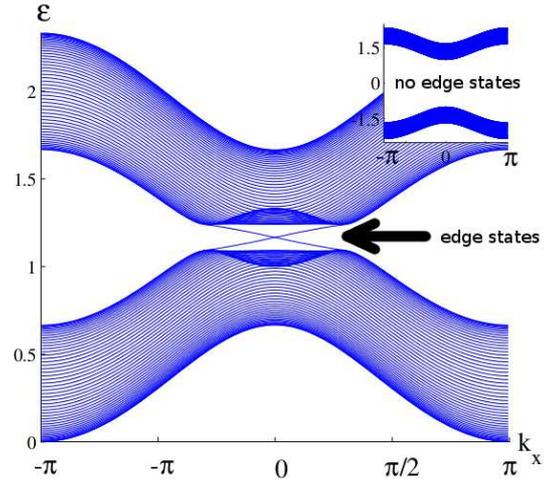}}
\caption{(Color online) The energy spectrum of a non-inverted HgTe/CdTe quantum well (inset) and the Floquet quasienergies in the presence of a linearly polarized perturbation (main panel) with 2 chiral edge modes, traversing the gap. Pictures taken from Ref. \cite{lindner}.}
\label{nphysfig}
\end{figure}

Starting from the trivial phase ($M/B<0$), and adding a linearly polarized perturbation $\mathcal{V}(\vk,t)= \vV(\vk) \cdot \bm{\sigma} \cos (\omega t)$, the authors of Ref. \cite{lindner} demonstrated that 
the effective bands are reshuffled (by resonant absorption processes) in such a way that the effective Hamiltonian is characterized by inverted effective bands, as shown in Fig. \ref{nphysfig}. Therefore the mechanism of band inversion, which is instrumental to the realization of both Chern and topological insulators \cite{KaneRMP:2011,QiRMP:2011}, can be realized by a suitable electromagnetic driving field in a broad range of photon energies and polarizations.

\section{Probing helical edge states by light \label{sectionQSH}}

We have seen that topological non-trivial band structures can in principle be created by shining light on a semimetal or on a trivial band insulator. Besides, it is natural to use photons to probe topological phases and their edge/surface states. We have studied a QSH state and its one-dimensional helical edge state in a circularly polarized radiation field \cite{Dora:2012}. Using Floquet theory, we have demonstrated that the photocurrent and the magnetization are ruled by the very same unit vector, whose winding number determines a topological invariant for the system. When increasing the radiation frequency, the edge state switches between a dissipationless quantized charge pumping behavior to a dissipative regime without quantization. This loss of quantization is caused by the Floquet band crossing which results in a "mixing'' of the topological invariants of individual bands. Our predictions could in principle be tested by experiments similar to those in graphene \cite{karch} and HgTe/CdTe quantum wells \cite{wittmann}.  

\subsection{Zeeman vs orbital coupling}

\begin{figure}[h!]
\begin{center}
{\includegraphics[width=5.0cm]{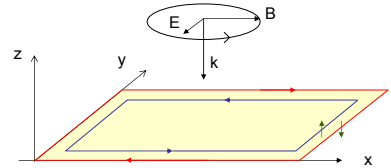}}
\caption{(Color online) The quantum spin-Hall insulator (light yellow rectangle) with its helical edge state (counterpropagating red/blue arrows) in a circularly polarized electromagnetic field with frequency $\omega$ and wave vector $k$. In the plane $z=0$ the rotating potential vector ${\bf A}(t)=A_0(-\sin\omega t,\cos\omega t)$ is perpendicular to the $S^z$ direction (vertical green arrows). Picture taken from Ref. \cite{Dora:2012}.} 
\end{center}
\label{deviceBB}
\end{figure}

We consider a QSH insulator located in the $xy$ plane and irradiated by a circularly polarized electromagnetic field with frequency $\omega$ (Fig. 3). The general Hamiltonian of the QSH edge reads in this setting:
\begin{gather}
H(t)=v_FS^z(p-eA_x(t))+g\left[S^+\exp(-i\omega t)+h.c.\right],
\label{hamilton}
\end{gather}
where $\bf S$ is the physical spin of the electron, $p$ the momentum along the one-dimensional channel and $v_F$ the Fermi velocity. It is assumed that the quantization axis of the QSH edge state is perpendicular to the plane $xy$. The circularly polarized radiation now acts on both the orbital motion through the vector potential $A_x(t)=-A_0 \sin \omega t$ and on the electron spin through the Zeeman coupling $g=g_{\rm eff}\mu_B B_0$, $g_{\rm eff}$ being the effective $g$-factor and $\mu_B$ the Bohr magneton. Nevertheless at high frequency, the orbital effect can be safely neglected according to a simple semi-classical argument: an electron travelling at speed $v_F$ in an electric field $E_0=A_0\omega=cB_0$ during a time $1/\omega$ picks up an energy $v_F e E_0/\omega$ from the vector potential which has to be compared to the smallest energy quantum it can absorb, $\hbar\omega$. Hence in the regime $v_F e E_0/\omega\ll \hbar \omega$, only the time-dependent Zeeman effect is effective and not the orbital effect. In contrast, in 2D systems the orbital effect is the dominant one \cite{Oka:2009,Calvo:2011,Hosur:2011,inoue,abergel,zhouwu,schmidt}.

\subsection{Topological invariant and photocurrent}

We now analyse the topological invariant describing the steady state of the QSH edge state in terms of the 
mapping, $(p,t) \rightarrow {\bf\hat d}_{\alpha,p}(t)=\Phi^+_\alpha(p,t){\bf S}\Phi_\alpha(p,t)=\alpha (g\cos(\omega t),g\sin(\omega t),v_Fp-\omega/2)/\lambda$, betwen the 1+1 dimensional
extended Brillouin zone in $(p,t)$ space 
and the unit sphere. Here,  $\alpha=\pm 1$ distinguishes between the Floquet bands,  $\lambda=\sqrt{g^2+(v_Fp-\omega/2)^2}$ and $\Phi_\alpha(p,t)$ is the Floquet wavefunction.

The sub-band Chern number, 
\begin{gather}
\mathcal{C}_\alpha=\frac 12 \sum_{\alpha,p}\int\limits_0^T dt {\bf \hat d}_{\alpha,p}(t)\cdot
 \left(\frac{\partial{\bf \hat d}_{\alpha,p}(t)}{\partial p}\times \frac{\partial{\bf \hat d}_{\alpha,p}(t)}{\partial t}\right),
\label{chern3d}
\end{gather}
counts the number of times the unit vector ${ \bf\hat d}_{\alpha,p}(t)$ wraps around the unit sphere \cite{KaneRMP:2011,lindner}, the summation being taken over occupied bands. 
The closely related photocurrent reads as
\begin{gather}
\langle j\rangle=\frac{e}{2\pi}|\omega|\mathcal{C}_{-}.
\label{photocurrent}
\end{gather}

The behavior of this topological invariant (and photocurrent) can be investigated as a function of the frequency of the driving field. At low frequency, $| \omega | < 4g$, and half filling, 
the ground state is the filled $\alpha=-1$ band and
\begin{gather}
\mathcal{C}_\alpha=-\int dp\frac{\alpha \textmd{sign}(\omega)v_Fg^2}{2\lambda^3}=-\alpha\textmd{sign}(\omega)~~
\label{chernwf}
\end{gather}
is quantized (Fig. 4, dashed line). By contrast, when $|\omega|>4g$, the two bands cross, and $\mathcal{C}_-$ is no longer quantized (Fig. 4), in analogy 
to the transfer of Chern numbers between equilibrium bands which touch:
\begin{gather}
\mathcal{C}_\alpha=-\alpha\textmd{sign}(\omega)\left(1-\sum_{s=\pm 1}s\frac{\sqrt{2\omega_s\omega}}{\omega}\right)
\label{chernnonquantized}
\end{gather}
vanishing as $\mathcal{C}_\alpha=-\alpha 2g/\omega$ for $g\ll|\omega|$. The quantized pumping rate is lost due to the crossing of Floquet subbands carrying opposite Chern number. It is remarkable
that the pumping rate is quantized over a broad range of frequencies, $|\omega| < 4g$, that exceeds the strict adiabatic limit. The robustness of the quantization for $|\omega| < 4g$ is a feature specific to our model.

To estimate the typical induced photocurrent, we recall that the $g\ll |\omega|$ regime is realized usually (and note that larger values of $\omega$ are beneficial for neglecting the vector potential), and for a radiation field with magnetic field strength of the order of $10^{-4}-10^{-5}$~T, this translates to a photocurrent of the order of $0.1-10$~pA, depending also on the effective $g$-factor values. These can be significantly enhanced ($g_{\rm eff}\approx 20-50$) for certain materials such as
HgTe/CdTe, HgSe or Bi$_2$Se$_3$.

\begin{figure}[h!]
\begin{center}
{\includegraphics[width=6cm]{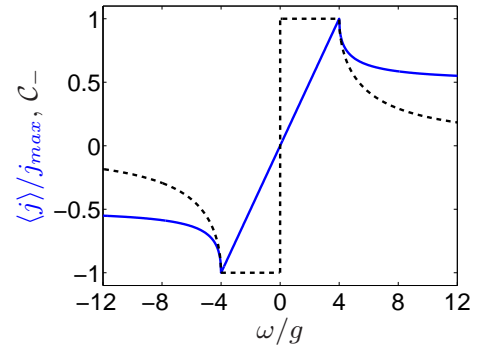}}
\caption{(Color online) The induced photocurrent (blue solid line) and the $\mathcal{C}_-$ Chern number (black dashed line) are shown as a function of $\omega/g$.
The latter becomes non-quantized when band touching occurs at $4g=|\omega|$.}
\end{center}
\label{jsz}
\end{figure}

The induced current can be detected in a contactless measurement. The photoinduced "rectified current" being of the order $\langle j\rangle=1$~pA, the corresponding magnetic field $B_{ind}=1$~pT (for a $1$ micron perimeter) is within the detectability limit of an ac SQUID \cite{Dora:2012}.


The topological properties of the edge state are reflected in the induced magnetization and photocurrent (Fig. 4). 
In the $|\omega|<4g$ regime, the current is obtained (upon restoring original units) as $\langle j \rangle=\textmd{sign}(\omega)e/T$, which 
tells us that the charge pumped within one cycle ($T$) is exactly the unit charge. The integer charge pumped across a 1D insulator in one 
period of an (adiabatic) cycle is a topological invariant that characterizes the cycle. This specific quantization of charge stems directly
from the quantized $\mathcal{C}_-=\textmd{sign}(\omega)$ in this regime, as was identified by Thouless \cite{thouless}. The current is dissipationless, protected 
by a photoinduced gap.
Though the current still satisfies $\langle j\rangle=e\mathcal{C}_-/T$ for $|\omega|>4g$, it is dissipative and no longer quantized due to the 
band touching, in analogy with the photovoltaic Hall effect \cite{Oka:2009} in graphene.

We also emphasize that using (neutral) atoms in optical traps, one can introduce a Zeeman term without 
the orbital counterpart or fabricate chiral edge states with spin quantized parallel to the momentum \cite{goldman}. 
Then the vector potential would also be absent and the full transition from dissipationless to dissipative charge pumping 
can be followed.

\section{Photocurrent induced within a 2D surface state \label{sectionPhotocurrent2D}}
We now discuss the potentialities of using circularly polarized light to probe the surface state (SS) of a three dimensional (3D) topological insulator, like Bi$_2$Se$_3$,  Bi$_2$Te$_3$, or strained HgTe  \cite{Hosur:2011,Ueda:2012}. 

{\it Detection of Berry curvature.}  In contrast to the previous section on 1D helical edge states, the orbital effect is the dominant coupling between light and electrons in a SS of a 3D TI. In its simplest version, the SS of 3D topological insulator consists in a single Dirac cone. Moreover each momentum state $\vk$ has a unique spin state (fixed by $\vk$). This spin-locking property allows a circularly polarized light to generate a DC rectified current at the surface of the insulator. Recently, Pavan Hosur pointed out that this so-called circular photogalvanic effect (CPE) can be used as a detection of a finite Berry curvature \cite{Hosur:2011}. Indeed in presence of the latter, there is a contribution to the photocurrent that grows linearly in time before being cut-off (by a scattering event) after some relaxation time. Since carrier relaxation times are expected to be quite large in the SS of a 3D TI, this current can overwhelm other contributions that are insensitive to the Berry curvature. An experimentally measurable DC current is predicted if the rotational symmetry of the SS state is broken by a strain or a static magnetic field. 

The CPE allows to detect the presence of hexagonal warping and spin tilting out-of the SS plane \cite{Fu:2009,Adroguer:2012}. A linearly dispersing Dirac cone with the spin winding in the plane will not give such a dominant current. In contrast, in presence of hexagonal warping, the spin is tilting out of the plane and the photocurrent will develop a current growing linearly in time. Besides, the CPE could also be used to probe the relaxation times in the SS of 3D Tis \cite{Hosur:2011}. 

{\it Pumping by the magnetization precession.} Ueda {\it et al.} have investigated the current dynamics within a SS of a 3D TI driven by the precession of the magnetization of an attached ferromagnet \cite{Ueda:2012}. Here the coupling mechanism between the oscillating magnetization and the SS carriers is the spin exchange coupling. A rectified DC current can be generated when the precession axis is within the SS plane. 

\section{Summary}

Recently it was shown that circularly polarized light might be used to open gaps in the electronic spectrum of graphene, in particular at the Dirac points. Furthermore such gaps would correspond to a topological mass of the Haldane type which can be detected through the associated Hall response. More generally a trivial insulator can be driven into a topological phase of matter by applying a proper time-periodic perturbation. When the static (non irradiated) system is already topological, photons are also useful to probe the properties and the robustness of the chiral/helical edge modes. Finally cold atomic vapors trapped in optical lattices provide very interesting routes to design synthetic gauge fields and induce topological phases \cite{Dalibard:2011}, either by using Raman resonances or by using periodic driving (shaking) of the lattice \cite{Hauke:2012}.  

\begin{acknowledgement}
We are grateful to Bernard Pla\c{c}ais and Beno\^it Dou\c{c}ot for helpful comments. JC acknowledges support from EU/FP7 under contract TEMSSOC and from ANR through project
2010-BLANC-041902 (ISOTOP). This research has been  supported by the Hungarian Scientific
Research Funds Nos.  K72613, K73361,
K101244, CNK80991,
T\'{A}MOP-4.2.1/B-09/1/KMR-2010-0002 and by the 
ERC Grant Nr. ERC-259374-Sylo.
\end{acknowledgement}

%
\bibliographystyle{pss}
\bibliography{HdrThesis.bib}

\providecommand{\WileyBibTextsc}{}
\let\textsc\WileyBibTextsc
\providecommand{\othercit}{}
\providecommand{\jr}[1]{#1}
\providecommand{\etal}{~et~al.}


\begin{thebibliography}{[10]}

\bibitem{Kane:2005a}
 \textsc{C.\,L. Kane} and  \textsc{E.\,J. Mele},
 \jr{Phys. Rev. Lett.} \textbf{95}, 226801 (2005).


\bibitem{Kane:2005b}
 \textsc{C.\,L. Kane} and  \textsc{E.\,J. Mele},
 \jr{Phys. Rev. Lett.} \textbf{95}, 146802 (2005).


\bibitem{Bernevig:2006}
 \textsc{B.~Bernevig},  \textsc{T.\,L. Hughes},  and  \textsc{S.\,C.
  Zhang},
 \jr{Science} \textbf{314}(5806), 1757 (2006).


\bibitem{KaneRMP:2011}
 \textsc{M.\,Z. Hasan} and  \textsc{C.\,L. Kane},
 \jr{Rev. Mod. Phys.} \textbf{82}(Nov), 3045--3067 (2010).


\bibitem{QiRMP:2011}
 \textsc{X.\,L. Qi} and  \textsc{S.\,C. Zhang},
 \jr{Rev. Mod. Phys.} \textbf{83}(Oct), 1057--1110 (2011).


\bibitem{Haldane:1988}
 \textsc{F.\,D.\,M. Haldane},
 \jr{Phys. Rev. Lett.} \textbf{61}, 2015 (1988).


\bibitem{Qi:2006}
 \textsc{X.\,L. Qi},  \textsc{Y.\,S. Wu},  and  \textsc{S.\,C. Zhang},
 \jr{Phys. Rev. B} \textbf{74}(Aug), 085308 (2006).


\bibitem{Tang:2011}
 \textsc{E.~Tang},  \textsc{J.\,W. Mei},  and  \textsc{X.\,G. Wen},
 \jr{Phys. Rev. Lett.} \textbf{106}, 236802 (2011).


\bibitem{Sun:2011}
 \textsc{K.~Sun},  \textsc{Z.~Gu},  \textsc{H.~Katsura},  and
  \textsc{S.~Das~Sarma},
 \jr{Phys. Rev. Lett.} \textbf{106}, 236803 (2011).


\bibitem{Neupert:2011}
 \textsc{T.~Neupert},  \textsc{L.~Santos},  \textsc{C.~Chamon},  and
  \textsc{C.~Mudry},
 \jr{Phys. Rev. Lett.} \textbf{106}, 236804 (2011).


\bibitem{Weeks:2010}
 \textsc{C.~Weeks} and  \textsc{M.~Franz},
 \jr{Phys. Rev. B} \textbf{82}(Aug), 085310 (2010).


\bibitem{guo}
 \textsc{H.\,M. Guo} and  \textsc{M.~Franz},
 \jr{Phys. Rev. B} \textbf{80}, 113102 (2009).


\bibitem{sun}
 \textsc{K.~Sun},  \textsc{H.~Yao},  \textsc{E.~Fradkin},  and  \textsc{S.\,A.
  Kivelson},
 \jr{Phys. Rev. Lett.} \textbf{103}, 046811 (2009).


\bibitem{Jiang:2011}
 \textsc{L.~Jiang},  \textsc{T.~Kitagawa},  \textsc{J.~Alicea},  \textsc{A.\,R.
  Akhmerov},  \textsc{D.~Pekker},  \textsc{G.~Refael},  \textsc{J.\,I. Cirac},
  \textsc{E.~Demler},  \textsc{M.\,D. Lukin},  and  \textsc{P.~Zoller},
 \jr{Phys. Rev. Lett.} \textbf{106}(Jun), 220402 (2011).


\bibitem{stanescu}
 \textsc{T.\,D. Stanescu},  \textsc{V.~Galitski},  \textsc{J.\,Y. Vaishnav},
  \textsc{C.\,W. Clark},  and  \textsc{S.~Das~Sarma},
 \jr{Phys. Rev. A} \textbf{79}(May), 053639 (2009).


\bibitem{goldman}
 \textsc{N.~Goldman},  \textsc{I.~Satija},  \textsc{P.~Nikolic},
  \textsc{A.~Bermudez},  \textsc{M.\,A. Martin-Delgado},
  \textsc{M.~Lewenstein},  and  \textsc{I.\,B. Spielman},
 \jr{Phys. Rev. Lett.} \textbf{105}(Dec), 255302 (2010).


\bibitem{QiPhysToday:2010}
 \textsc{X.\,L. Qi} and  \textsc{S.\,C. Zhang},
 \jr{Phys. Today} \textbf{63}, 33 (2010).


\bibitem{Konig:2007}
 \textsc{M.~K{\"o}nig},  \textsc{S.~Wiedmann},  \textsc{C.~Br{\"u}ne},
  \textsc{A.~Roth},  \textsc{H.~Buhmann},  \textsc{L.\,W. Molenkamp},
  \textsc{X.\,L. Qi},  and  \textsc{S.\,C. Zhang},
 \jr{Science} \textbf{318}(5851), 766--770 (2007).


\bibitem{Roth:2009}
 \textsc{A.~Roth},  \textsc{C.~Brune},  \textsc{H.~Buhmann},  \textsc{L.\,W.
  Molenkamp},  \textsc{J.~Maciejko},  \textsc{X.\,L. Qi},  and  \textsc{S.\,C.
  Zhang},
 \jr{Science} \textbf{325}(5938), 294--297 (2009).


\bibitem{kitagawa}
 \textsc{T.~Kitagawa},  \textsc{E.~Berg},  \textsc{M.~Rudner},  and
  \textsc{E.~Demler},
 \jr{Phys. Rev. B} \textbf{82}(Dec), 235114 (2010).


\bibitem{Ryu:2008}
 \textsc{A.\,P. Schnyder},  \textsc{S.~Ryu},  \textsc{A.~Furusaki},  and
  \textsc{A.\,W.\,W. Ludwig},
 \jr{Phys. Rev. B} \textbf{78}(Nov), 195125 (2008).


\bibitem{Hughes:2008}
 \textsc{X.\,L. Qi},  \textsc{T.\,L. Hughes},  and  \textsc{S.\,C.
  Zhang},
 \jr{Phys. Rev. B} \textbf{78}(Nov), 195424 (2008).


\bibitem{Kitaev:2009}
 \textsc{A.~Kitaev},
 \jr{Advances in Theoretical Physics: Landau Memorial Conference Chernogolovska
  (Russia) 22-26 June 2008, AIP Conf. Proc.} \textbf{1134}, 22 (2009).


\bibitem{lindner}
 \textsc{N.\,H. Lindner},  \textsc{G.~Refael},  and
  \textsc{V.~Galitski},
 \jr{Nat. Phys.} \textbf{7}, 490 (2011),
arXiv:1008.1792.


\bibitem{Oka:2009}
 \textsc{T.~Oka} and  \textsc{H.~Aoki},
 \jr{Phys. Rev. B} \textbf{79}, 081406 (2009).


\bibitem{Calvo:2011}
 \textsc{H.~Calvo},  \textsc{H.~Pastawski},  \textsc{S.~Roche},  and
  \textsc{L.~Torres},
 \jr{Appl. Phys. Lett.} \textbf{98}, 232103 (2011).


\bibitem{Kitagawa:2011}
 \textsc{T.~Kitagawa},  \textsc{T.~Oka},  \textsc{A.~Brataas},  \textsc{L.~Fu},
   and  \textsc{E.~Demler},
 \jr{Phys. Rev. B} \textbf{84}, 235108 (2011),
arXiv:1104.4636.


\bibitem{Morell:2012}
 \textsc{E.~Morell} and  \textsc{L.~Torres},
 \jr{Phys. Rev. B} \textbf{86}, 125449 (2012).


\bibitem{GuPRL:2011}
 \textsc{Z.~Gu},  \textsc{H.\,A. Fertig},  \textsc{D.\,P. Arovas},  and
  \textsc{A.~Auerbach},
 \jr{Phys. Rev. Lett.} \textbf{107}(Nov), 216601 (2011).


\bibitem{Dora:2012}
 \textsc{B.~D{\'o}ra},  \textsc{J.~Cayssol},  \textsc{F.~Simon},  and
  \textsc{R.~Moessner},
 \jr{Phys. Rev. Lett.} \textbf{108}, 056602 (2012).


\bibitem{Hosur:2011}
 \textsc{P.~Hosur},
 \jr{Phys. Rev. B} \textbf{83}(Jan), 035309 (2011).


\bibitem{Ueda:2012}
 \textsc{H.\,T. Ueda},  \textsc{A.~Takeuchi},  \textsc{G.~Tatara},  and
  \textsc{T.~Yokoyama},
 \jr{Phys. Rev. B} \textbf{85}(Mar), 115110 (2012).


\bibitem{sambe}
 \textsc{H.~Sambe},
 \jr{Phys. Rev. A} \textbf{7}(Jun), 2203--2213 (1973).


\othercit
\bibitem{dittrich}
 \textsc{T.~Dittrich},  \textsc{P.~Hanggi},  \textsc{G.\,L. Ingold},
  \textsc{B.~Kramer},  \textsc{G.~Schon},  and  \textsc{W.~Zwerger} (eds.),
Quantum Transport and Dissipation (Wiley-WCH, Weinheim, 1998).


\bibitem{Asboth:2012}
 \textsc{J.\,K. Asb\'oth},
 \jr{Phys. Rev. B} \textbf{86}(Nov), 195414 (2012).


\bibitem{Thouless1982}
 \textsc{D.\,J. Thouless},  \textsc{M.~Kohmoto},  \textsc{M.\,P. Nightingale},
  and  \textsc{M.~den Nijs}\textbf{49}(6), 405 (1982).


\bibitem{inoue}
 \textsc{J.\,i. Inoue} and  \textsc{A.~Tanaka},
 \jr{Phys. Rev. Lett.} \textbf{105}, 017401 (2010).


\bibitem{Liu:2008}
 \textsc{C.~Liu},  \textsc{T.\,L. Hughes},  \textsc{X.\,L. Qi},
  \textsc{K.~Wang},  and  \textsc{S.\,C. Zhang},
 \jr{Phys. Rev. Lett.} \textbf{100}(Jun), 236601 (2008).


\bibitem{Knez:2011}
 \textsc{I.~Knez},  \textsc{R.\,R. Du},  and  \textsc{G.~Sullivan},
 \jr{Phys. Rev. Lett.} \textbf{107}(Sep), 136603 (2011).


\bibitem{karch}
 \textsc{J.~Karch},  \textsc{P.~Olbrich},  \textsc{M.~Schmalzbauer},
  \textsc{C.~Zoth},  \textsc{C.~Brinsteiner},  \textsc{M.~Fehrenbacher},
  \textsc{U.~Wurstbauer},  \textsc{M.\,M. Glazov},  \textsc{S.\,A. Tarasenko},
  \textsc{E.\,L. Ivchenko},  \textsc{D.~Weiss},  \textsc{J.~Eroms},
  \textsc{R.~Yakimova},  \textsc{S.~Lara-Avila},  \textsc{S.~Kubatkin},  and
  \textsc{S.\,D. Ganichev},
 \jr{Phys. Rev. Lett.} \textbf{105}(Nov), 227402 (2010).


\bibitem{wittmann}
 \textsc{B.~Wittmann},  \textsc{S.\,N. Danilov},  \textsc{V.~Bel'kov},
  \textsc{S.\,A. Tarasenko},  \textsc{E.\,G. Novik},  \textsc{H.~Buhmann},
  \textsc{C.~Br{\"u}ne},  \textsc{L.\,W. Molenkamp},  \textsc{Z.\,D. Kvon},
  \textsc{N.\,N. Mikhailov},  \textsc{S.\,A. Dvoretsky},  \textsc{N.\,Q. Vinh},
   \textsc{A.\,F.\,G. {van der Meer}},  \textsc{B.~Murdin},  and
  \textsc{S.\,D. Ganichev},
 \jr{Semicond. Sci. Technol.} \textbf{25}, 095005 (2010).


\bibitem{abergel}
 \textsc{D.\,S.\,L. Abergel} and  \textsc{T.~Chakraborty},
 \jr{Nanotechnology} \textbf{22}, 015203 (2011).


\bibitem{zhouwu}
 \textsc{Y.~Zhou} and  \textsc{M.\,W. Wu},
 \jr{Phys. Rev. B} \textbf{83}(Jun), 245436 (2011).


\bibitem{schmidt}
 \textsc{M.\,J. Schmidt},  \textsc{E.\,G. Novik},  \textsc{M.~Kindermann},  and
   \textsc{B.~Trauzettel},
 \jr{Phys. Rev. B} \textbf{79}(Jun), 241306 (2009).


\bibitem{thouless}
 \textsc{D.\,J. Thouless},
 \jr{Phys. Rev. B} \textbf{27}(May), 6083--6087 (1983).


\bibitem{Fu:2009}
 \textsc{L.~Fu},
 \jr{Phys. Rev. Lett.} \textbf{103}, 266801 (2009).


\bibitem{Adroguer:2012}
 \textsc{P.~Adroguer},  \textsc{D.~Carpentier},  \textsc{J.~Cayssol},  and
  \textsc{E.~Orignac},
 \jr{New J. Phys.} \textbf{14}, 103027 (2012).


\bibitem{Dalibard:2011}
 \textsc{J.~Dalibard},  \textsc{F.~Gerbier},
  \textsc{G.~Juzeli\ifmmode\,\bar{u}\else \={u}\fi{}nas},  and
  \textsc{P.~\"Ohberg},
 \jr{Rev. Mod. Phys.} \textbf{83}(Nov), 1523--1543 (2011).


\bibitem{Hauke:2012}
 \textsc{P.~Hauke},  \textsc{O.~Tieleman},  \textsc{A.~Celi},
  \textsc{C.~\"Olschl\"ager},  \textsc{J.~Simonet},  \textsc{J.~Struck},
  \textsc{M.~Weinberg},  \textsc{P.~Windpassinger},  \textsc{K.~Sengstock},
  \textsc{M.~Lewenstein},  and  \textsc{A.~Eckardt},
 \jr{Phys. Rev. Lett.} \textbf{109}(Oct), 145301 (2012).


\end{thebibliography}
%


\end{document}